\documentclass[%
twocolumn,
amsmath,amssymb,
aps,
superscriptaddress,
]{revtex4-2}

\usepackage{url}
\usepackage{graphicx}
\graphicspath{{.}{figs/}{../figs/}}
\usepackage{afterpage}
\usepackage{siunitx}
\DeclareSIUnit\angstrom{\text {Å}}
\usepackage{dcolumn}
\usepackage{bm}
\usepackage{tikz}
\usetikzlibrary{positioning,intersections,angles,quotes}
\usepackage[caption=false]{subfig}
\usepackage{booktabs} 
\usepackage{comment}
\usepackage{physics}
\usepackage{mathtools}

\usepackage[colorlinks=true,citecolor=blue,linkcolor=magenta]{hyperref}

\usepackage{longtable}
\usepackage{booktabs}

\begin{document}

\title{Ising model formulation for highly accurate topological color codes decoding}

\author{Yugo Takada}
\email{u751105k@ecs.osaka-u.ac.jp}
\affiliation{%
  Graduate School of Engineering Science, Osaka University, 1-3 Machikaneyama, Toyonaka, Osaka 560-8531, Japan
}
\author{Yusaku Takeuchi}
\email{u256654e@ecs.osaka-u.ac.jp}
\affiliation{%
  Graduate School of Engineering Science, Osaka University, 1-3 Machikaneyama, Toyonaka, Osaka 560-8531, Japan
}%
\author{Keisuke Fujii}%
\email{fujii@qc.ee.es.osaka-u.ac.jp}
\affiliation{%
  Graduate School of Engineering Science, Osaka University, 1-3 Machikaneyama, Toyonaka, Osaka 560-8531, Japan
}%
\affiliation{%
  Center for Quantum Information and Quantum Biology,
  Osaka University, 1-2 Machikaneyama, Toyonaka 560-0043, Japan
}%
\affiliation{%
  RIKEN Center for Quantum Computing (RQC),
  Hirosawa 2-1, Wako, Saitama 351-0198, Japan
}%
%
\date{\today}
\begin{abstract}
  Quantum error correction is an essential ingredient for reliable quantum computation for theoretically provable quantum speedup.
  Topological color codes, one of the quantum error correction codes, have an advantage against the surface codes in that all Clifford gates can be implemented transversally.
  However, the hardness of decoding makes the color codes not suitable as the best candidate for experimentally feasible implementation of quantum error correction.
  Here we propose an Ising model formulation that enables highly accurate decoding of the color codes. 
  In this formulation, we map stabilizer operators to classical spin variables to represent an error satisfying the syndrome.
  Then we construct an Ising Hamiltonian that counts the number of errors and formulate the decoding problem as an energy minimization problem of an Ising Hamiltonian, which is solved by simulated annealing.
  In numerical simulations on the (4.8.8) lattice, we find an error threshold of 10.36(5)\% for bit-flip noise model, 
  18.47(5)\% for depolarizing noise model, and 2.90(4)\% for phenomenological noise model (bit-flip error is located on each of data and measurement qubits), 
  all of which are higher than the thresholds of existing efficient decoding algorithms.
  Furthermore, we verify that the achieved logical error rates are almost optimal 
  in the sense that they are almost the same as those obtained by exact optimizations by CPLEX with smaller decoding time in many cases.
  Since the decoding process has been a bottleneck for performance analysis, 
  the proposed decoding method is useful for further exploration of the possibility of the topological color codes.
\end{abstract}

\maketitle

\section{Introduction}
In recent years, there has been remarkable progress in the development of quantum computers,
with the realization of quantum computers of tens to hundreds of qubits. 
While entering the realm of hard-to-simulate areas for classical computers~\cite{arute2019quantum,zhong2020quantum,madsen2022quantum,ball2021first}, demonstrating quantum advantage in meaningful problems remains challenging. 
This is due to the current high noise level of quantum computers, which hinders the execution of complex quantum algorithms,
such as Shor's factorization~\cite{shor}, linear system solver~\cite{hhl}, and quantum phase estimation~\cite{phase_estimation}. 
To solve this problem, it is essential to realize a fault-tolerant quantum computer, which can perform computation while protecting quantum information from errors through quantum error correction (QEC)~\cite{shor1995scheme}. 

The surface codes \cite{surface}, one of the topological codes \cite{topological},
are thought to be one of the most promising approaches for the experimental implementation of QEC
due to their simple structure, making physical implementation easier, with a relatively high error threshold. 
Currently, experimental demonstrations of QEC on a single logical qubit level of the surface code are ongoing \cite{,krinner2022realizing,zhao2022realization,acharya2022suppressing}.
Nevertheless, surface codes have a drawback as QEC codes. Specifically, only $X$, $Z$, and CNOT gates can be implemented transversally among the Clifford gates \cite{surface_transversal}. 
Consequently, a special treatment is required to implement $H$ and $S$ gate fault-tolerantly \cite{lattice_surgery1,lattice_surgery2,lattice_surgery3}, resulting in additional overhead.

The QEC codes that resolve this drawback are the color codes \cite{color1,color2}. 
A color code has an advantage against the surface codes in that all Clifford gates can be implemented transversally,
due to its high symmetry of the stabilizer operators \cite{landahl}. 
Despite this advantage, color codes are currently not the mainstream
of experimental implementations of QEC due to the disadvantages,
the difficulty of decoding and its low error threshold under the circuit-level noise model. 
For the surface codes, there is a known decoding algorithm, minimum-weight perfect matching algorithm (MWPM) \cite{mwpm}, which can be executed in polynomial time and with high accuracy. 
On the other hand, in the case of color codes, such a good decoding algorithm is still missing. 
The optimal threshold of a color code on the (4.8.8) lattice is estimated 
to be 10.9\% \cite{x_threshold} under the bit-flip noise model by a Monte Carlo simulation of the 
corresponding statistical mechanical model.
However, the performances of efficient decoders, such as the renormalization group decoder \cite{renormalize}, the restriction decoder with MWPM \cite{unionfind}, and the restriction decoder with union-find \cite{unionfind}, resulting in a threshold of 8.7\%, 10.2\%, and 9.8\% (analytical estimate), respectively. Although the integer program decoder \cite{landahl} exhibits good accuracy with a threshold of 10.6\%, 
its decoding process incurs exponential computational time. 
Regarding the depolarizing noise model, 
the optimal threshold is estimated to be 18.9\% \cite{xyz_threshold}. 
Nonetheless, the neural-network decoder \cite{neural} has a comparatively low accuracy of 17.5\%. 
Moreover, for the phenomenological noise model, 
which takes into account errors that occur during syndrome measurement, 
prior studies have shown that the graph matching decoder has a poor accuracy of 2.08\% \cite{graphmatching}.
Also, the integer program decoder \cite{landahl} has better accuracy at 3.05\%, but requires exponential time for decoding.
Although it has been demonstrated that tensor network decoders can achieve a relatively high threshold for the bit-flip and depolarizing noise model on the (6.6.6) color codes \cite{PhysRevX.9.041031,chubb2021general}, it is unclear how high of a threshold value can be obtained for the phenomenological noise model.
The threshold for (6.6.6) color codes under phenomenological noise was estimated to be 4.8(3)\% \cite{andrist2011tricolored,PhysRevA.94.012318} by mapping the problem onto a statistical–mechanical three-dimensional disordered Ising lattice gauge theory, but it is not the value obtained by decoders.
The difficulty in decoding implies a challenge in evaluating its performance, 
thus obstructing the exploration of methods to overcome the issue of a low error threshold
under the circuit-level noise model, which is another drawback of the color codes.

In this paper, we propose a Monte Carlo type decoder for color codes. 
Monte Carlo type methods have been applied for decoding topological codes so far~\cite{fujii2014measurement,fujisaki2022practical}.
However, in these approaches, an Ising spin variable is assigned for each error, and then syndrome constraints are imposed as penalty terms, leading to relatively low accuracy of the error correction.
In the proposed scheme, we instead assign an Ising spin variable for each stabilizer operator.
Then we construct an Ising Hamiltonian that counts the number of errors via the multi-body Ising interactions
and formulates a decoding algorithm by minimizing the energy of the Ising Hamiltonian, 
which we solve by using simulated annealing (SA) \cite{sa}.
Specifically, when syndrome measurement is perfect, assigning an Ising spin variable for each stabilizer generator allows us to represent all error patterns. However, when errors are also located on measurement qubits, it becomes not so straightforward. In this situation, we introduce {\it time-like stabilizer generators} and assign an Ising spin variable for each of these. As a result, we can represent all patterns of both data errors and measurement errors and then construct an Ising Hamiltonian that counts the number of errors appropriately.
The syndrome constraints are imposed on the initial spin configuration
and hence satisfied throughout the optimization process.

We perform extensive numerical experiments to examine the performance of the proposed decoder
using an open source SA solver and compare its performance with CPLEX \cite{cplex},
which solves the same problems exactly via integer programming.
We find the error threshold of 10.36(5)\% for the bit-flip noise model, 
which is almost equivalent to that obtained by the integer program decoder with the highest accuracy in previous studies~\cite{landahl}. 
For the depolarizing noise model, 
we find 18.47(5)\%, which is higher than the neural-network decoder~\cite{neural}. 
For the phenomenological noise model, 
where only bit-flip errors occur as data errors and errors are also introduced on the measured syndrome,
we find 2.90(4)\%, which is again almost equivalent to that of the integer program decoder. 
While numerical experiments for the circuit-level noise model were not conducted, this decoder can also be applied straightforwardly for the circuit-level noise model.

In terms of decoding time, 
the proposed method is faster than CPLEX in the cases of the bit-flip noise model and depolarizing noise model. 
On the other hand, for the phenomenological noise model, 
the proposed method is faster than CPLEX only for a small code distance.
However, we should note that SA employed here can be further optimized for the Ising models formulated here. 
Furthermore, SA holds the advantage of enabling parallel computation. 
Therefore, we believe that optimizing and parallelizing SA for this task could lead to decoding with high accuracy while faster than other decoders that achieve comparable accuracy for a wider class of QEC code.

The rest of the paper is organized as follows.
In Sec \ref{preliminary}, we first introduce the definition and characteristics of color codes. 
Then, we discuss the problems of one of the existing decoders that is closely related to our method, which formulates the decoding problem 
as a combinatorial optimization problem by assigning variables to each error.
In Sec \ref{Decoding algorithm for code capacity noise}, we provide a detailed description of our decoding formulation for two types of code capacity noise models: the bit-flip noise model and the depolarizing noise model.
In Sec \ref{Decoding algorithm for phenomenological noise}, we describe the decoding formulation for the phenomenological noise model. 
Firstly, we introduce an existing decoding method that formulates the decoding problem as a combinatorial optimization problem by assigning variables to each error, similar to Sec \ref{preliminary} for the bit-flip noise model.
Next, we provide a detailed explanation of our decoding formulation in this noise model.
In Sec \ref{Numerical experiments}, we provide details of the numerical experiments performed and show the logical error rates achieved by our method through Monte Carlo simulations for each noise model. 
We also report the decoding time measured by the computer and provide a comparison and discussion with CPLEX.
Then, Sec \ref{conclusion} is devoted to a conclusion.

\section{Preliminary}
\label{preliminary}
\subsection{Color codes}
\begin{figure}[t]
  \begin{center}
        \includegraphics[width=0.9\linewidth]{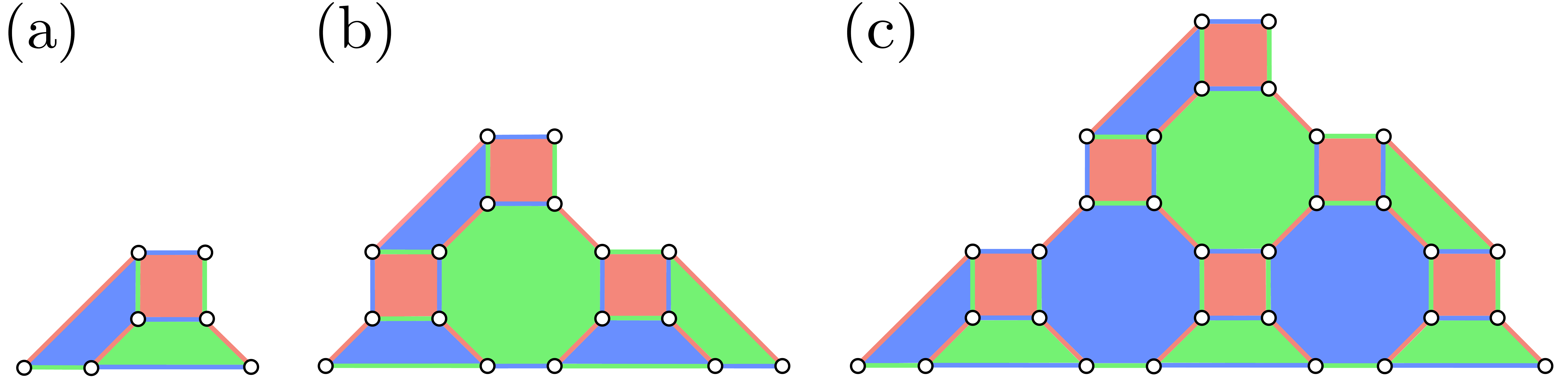}
  \end{center}
  \caption{\textbf{(4.8.8) color codes.} (a) $d=3$. (b) $d=5$. (c) $d=7$. $d$ denotes code distance.}
  \label{fig:lattice}
\end{figure}

Color codes are topological codes defined on a lattice where each vertex is incident to three edges and adjacent faces are colored with three different colors \cite{color1}, as shown in Fig.~\ref{fig:lattice}.
Also, the edges of the lattice are colored with three different colors.
Qubits are placed on each vertex of the lattice. 
For each face, the $Z$- and $X$-stabilizer generators are defined as the tensor product of Pauli $Z$ and $X$ operators acting on each qubit located at the vertices included in the face:

\begin{equation}
\label{colorzsta}
G_{Zf}:=\prod_{v \in f} Z_v,
\end{equation}
\begin{equation}
\label{colorxsta}
G_{Xf}:=\prod_{v \in f} X_v.
\end{equation}
The code state is defined as the simultaneous +1 eigenspace of stabilizer generators.
Specifically, we here use the color codes defined on the (4.8.8) lattice with open boundaries, depicted in Fig.~\ref{fig:lattice}. 
The (4.8.8) lattice is a semi-regular lattice where a square and two octagons meet at every vertex. 
Similar notations apply to other lattices as well, for example, the lattice denoted as (6.6.6) lattice is a regular lattice where three hexagons meet at every vertex.
In terms of transversal gates, a method is proposed to transversally implement all Clifford gates on any 2D color code \cite{PhysRevA.91.032330}.
However, the (4.8.8) color codes are the only 2D color codes that can implement all Clifford gates transversally in a simple way \cite{landahl}.

\subsection{Decoding and combinatorial optimization}
For simplicity, let us first consider decoding under the bit-flip noise model. 
Bit-flip noise model is a noise model in which the Pauli $X$ operator acts on each physical qubit with probability $p$. This noise model is written by the following map:

\begin{equation}
\mathcal{E}(\rho)=(1-p)\rho +p X \rho X.
\end{equation}
The problem of finding the error configuration that minimizes the number of errors satisfying the syndrome is formulated as the integer programming problem:

\begin{equation}
\label{min}
\mathrm{min} \quad \sum_{i} x_i,
\end{equation}

\begin{equation}
\label{sto}
\mathrm{s.t.} \quad \bigoplus_{i \in f} x_i =s_f \quad \forall f,
\end{equation}
where $x_i \in \{0,1 \}$ denotes a binary variable representing the error on the $i$-th qubit; $x_i=1$ ($=0$) represents the presence (absence) of error. 
$s_f \in \{ 0,1\}$ is the syndrome value of the $Z$-stabilizer generator defined on each face $f$. 
The minimum distance decoding is executed by exactly solving the integer programming problem 
represented by Eqs.~(\ref{min}) and (\ref{sto}) \cite{landahl}. 
While it has not been known whether or not the problem of decoding the color codes is NP-hard, in general, decoding problem of QEC codes formulated as an integer programming problem is NP-hard \cite{PhysRevA.83.052331}, making it unrealistic when the system size is large.

In this paper, we propose a decoding algorithm using SA, which is a heuristic approach for solving combinatorial optimization problems. 
For example,
the integer programming problem represented by Eqs.~(\ref{min}) and (\ref{sto}) is rewritten 
as an unconstrained optimization problem by introducing a penalty term~\cite{fujii2014measurement,fujisaki2022practical}:
\begin{equation}
\label{pena}
\mathrm{min} \quad \sum_{i} x_i +\lambda \left( \bigoplus_{i \in f} x_i \oplus s_f \right),
\end{equation}
where $\lambda$ is a hyperparameter referred to as the penalty coefficient and its value must be manually set beforehand. 
The difficulty in adjusting this penalty coefficient results in poor accuracy.
To achieve high accuracy, we avoid this formulation and 
map the decoding problem to an unconstrained optimization problem, where no penalty term for the syndrome constraints is introduced.

\section{Decoding algorithm for code capacity noise}
\label{Decoding algorithm for code capacity noise}
\subsection{Mapping decoding problem to an Ising model}
Here, we adopt a different approach to formulate the decoding problem under bit-flip noise as an integer programming problem.
Let us use the fact that the Pauli error $E$ can be decomposed as
\begin{equation}
  \label{tgl}
E=T(S)GL
\end{equation}
as shown in Ref.~\cite{poulin}. Here, $T(S)$ is a pure error, which is a Pauli operator that returns a quantum state with the syndrome $S$ to the code space. $G$ is a stabilizer operator and $L$ is a logical operator. 
The proof that the Pauli error can be decomposed as shown in Eq.~(\ref{tgl}) is as follows.
$T(S)E$ is an operator that acts within the code space, and operators acting within the code space can always be expressed as a product of a stabilizer operator and a logical operator. Therefore, for certain $G$ and $L$, the relation $T(S)E=GL$ always holds and it is shown that the Pauli error $E$ can be decomposed as Eq.~(\ref{tgl}).
Instead of considering the error $E$ as a variable, we consider the stabilizer operator $G$ in this decomposition formula as a variable. 
Thereby, the error $E$ represented in this manner can always satisfy the syndrome $S$ for an arbitrary stabilizer $G$.

\begin{figure}[b]
  \begin{center}
        \includegraphics[width=0.9\linewidth]{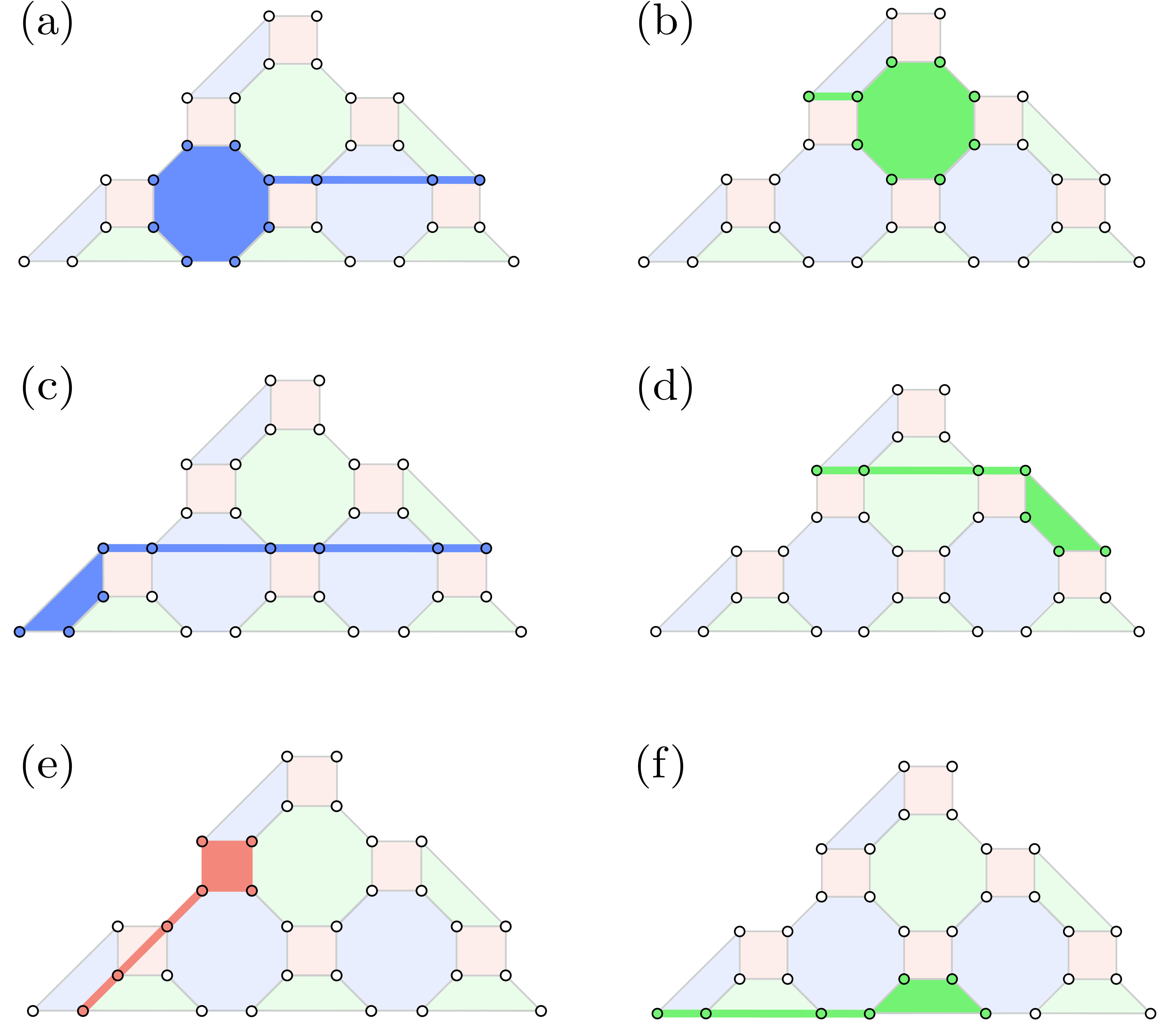}
  \end{center}
  \caption{\textbf{Definition of $T(S)$ for $d=7$ color codes.} The figure shows the correspondence between certain faces and the error chains associated with the construction of $T(S)$.
(a) The blue octagon. (b) The green octagon. (c) The blue trapezoid other than the bottom. (d) The green trapezoid other than the bottom. (e) The red square. (f) The bottom green trapezoid.}
  \label{fig:ts}
\end{figure}

While the definition of $T(S)$ is not unique, it should be defined in a systematic way for different code distances to simplify the implementation of the decoding algorithm. 
Specifically, we define $T(S)$ as follows (see Fig.~\ref{fig:ts}). 
First, we define an error chain that only anti-commutes with 
the stabilizer generator on each type of face:
\begin{itemize}
\item The blue face releases error chains from its vertex on the right side  through blue edges to the right boundary. ((a), (c) of Fig.~\ref{fig:ts})
\item The green face releases error chains from its vertex on the left side through green edges to the left boundary. ((b), (d), (f) of Fig.~\ref{fig:ts})
\item The red square releases error chains from its left lower vertex through red edges to the left bottom, until the bottom edge. ((e) of Fig.~\ref{fig:ts})
\end{itemize}
Then, these error operators, which are defined uniquely for each face operator, 
are multiplied together for $f$ with the syndrome value $s_f=1$ to define $T(S)$ for a given syndrome $S$.
An example of $T(S)$ for $d=7$ color codes is shown in Fig.~\ref{fig:ts}.
Although there are other strategies to choose $T(S)$ \cite{PhysRevLett.109.160503}, these are out of scope in this work because putting in significant effort into determining $T(S)$ can make it difficult to evaluate the performance of SA itself.

In order to define a binary optimization problem, 
we rewrite the operators $T(S),\ G,\ L$ in terms of binary variables $\{ t_i(S)\}, \{ g_i\}, l$ as follows:
\begin{equation}
T(S)=\prod_{i} X_i^{t_i(S)},
\end{equation}
\begin{equation}
G=\prod_{i} G_i^{g_i},
\end{equation}
\begin{equation}
L= L^l_X,
\end{equation}
where $t_i(S)$ is the bit representing the action of $T(S)$ on the $i$-th qubit,
$g_i$ is a binary variable indicating whether or not the $i$-th $X$-stabilizer generator $G_i$ is included,
and $l$ is a binary variable indicating whether or not the $X$-logical operator $L_X$ is included.
Note that by changing $l$ and $g_i$, we can represent any operator that commutes with the $Z$-stabilizer and hence does not change the syndrome value. 
On the other hand, we can write the error $E$ in terms of the binary variables $\{x_i\}$ as 
\begin{align}
    E= \prod _i X_i ^{x_i}.
\end{align}
By combining these, 
we have 
\begin{equation}
\label{xi}
x_i=t_i(S) \oplus \left( \bigoplus_{j \in B_i} g_j \right) \oplus (l_i \cdot l),
\end{equation}
where $B_i$ denotes the set of indices of the stabilizer generators acting on the $i$-th qubit, 
and $l_i$ is a binary variable by which the logical operator $L_X$ is given by $L_X = \prod_i X_i^{l_i}$.
By changing the binary variables to Ising spin variables via $\sigma_j := 1 - 2g_j$ and $J_i := (1-2 l_i \cdot l)(1-2t_i(S))$ in Eq.~(\ref{xi}), 
we obtain the equation
\begin{equation}
x_i=\frac{1-J_i \prod_{j \in B_i} \sigma_j}{2},
\end{equation}
so the number of errors $\sum_i x_i$ is expressed as
\begin{equation}
\sum_i \frac{1-J_i \prod_{j \in B_i} \sigma_j}{2} =\frac{1}{2} \left( N- \sum_{i} J_i \prod_{j \in B_i} \sigma_j \right),
\end{equation}
where $N$ is the number of qubits. Accordingly, we can rewrite the number of errors $\sum_i x_i$ as an Ising Hamiltonian:
\begin{equation}
\label{hubo}
H_l =-\sum_{i} J_i \prod_{j \in B_i} \sigma_j,
\end{equation}
where constant factor and term are omitted since they are not important. The Hamiltonian $H_l$ corresponds to the three-body Ising model.
By minimizing the Hamiltonian $H_l$ in Eq.~(\ref{hubo}), it is possible to obtain an error configuration that minimizes the number of errors. 
It should be noted that $H_l$ depends on $l$, and it is necessary to consider which case, $l$ being 0 or 1, yields the error configuration with the minimum number of errors. 
The error configuration that minimizes the Hamiltonian is determined for each case with $l=0,1$, and the error configuration achieving smaller energy between the two is the one that provides the minimum number of errors.
Thus, we can decode for the bit-flip noise model by solving 
\begin{equation}
\label{hubo_min}
\min _{l}  \left( \mathrm{min} \; H_l \right).
\end{equation}
Therefore, the decoding problem can be formulated as an energy minimization problem of an Ising Hamiltonian.
By solving Eq.~(\ref{hubo_min}) using SA, the optimized $g_j$ resulting in the minimum number of errors are obtained.

\subsection{Decoder for the depolarizing noise model}
So far we have only considered the bit-flip noise model.
Next, we extend the previous argument to a more general noise model, the depolarizing noise model.
The depolarizing noise model is a noise model in which the Pauli $X,\ Y$, and $Z$ operators act on each physical qubit with the probability $p_x=p_y=p_z=p/3$:
\begin{equation}
  \mathcal{E}(\rho)=(1-p)\rho +\frac{p}{3}(X \rho X +Y \rho Y +Z \rho Z  ).
\end{equation}
In the case of the depolarizing noise model, we can consider $Z$ errors in exactly the same way as $X$ errors,
since $X$-stabilizer and $Z$-stabilizer are symmetric on the color codes.
With respect to $Y$ errors, if $X$ and $Z$ errors occur simultaneously on a qubit, such an event should be counted as a $Y$ error.
Therefore, the total number of $X$, $Y$, and $Z$ errors is given by 
\begin{equation}
\label{num_error_depolar}
\sum_{i} x_i +\sum_{i} z_i -\sum_{i} x_i z_i,
\end{equation}
where $z_i \in \{0,1 \}$ is a binary variable that represents the presence or absence of the $Z$ error in the $i$-th qubit.
The mapping between $T(S),\ G,\ L$ and binary variables are given as follows:
\begin{equation}
T(S)=\prod_{i} X_i^{t^X_i(S)} \prod_{j}Z_j^{t^Z_j(S)},
\end{equation}
\begin{equation}
G=\prod_{i} G_{Xi}^{g^X_i} \prod_{j}G_{Zj}^{g^Z_j},
\end{equation}
\begin{equation}
L=L^{l_X}_X L^{l_Z}_Z,
\end{equation}
where $t^X_i(S),\ t^Z_i(S)$ are bits representing the action of $T(S)$ on the $i$-th qubit with respect to $X$ and $Z$, respectively. 
$g^X_i$ and $ g^Z_i$ are binary variables representing whether or not the $X$- and $Z$-stabilizer generators $G_{Xi}$ and $G_{Zi}$ are included, respectively. 
$l_X$ and $l_Z$ are binary variables representing whether or not the $X$- and $Z$-logical operators $L_X$ and  $L_Z$ are included, respectively.
Similar to the discussion in the case of bit-flip noise, we can write the error $E$ in terms of binary variables $\{ x_i \}$, $\{ z_i \}$ as
\begin{equation}
    E= \prod _i X_i ^{x_i} \prod _j Z_j ^{z_j}.
\end{equation}
By combining these, $x_i$ and $z_i$ are represented as
\begin{equation}
\label{xi_y}
x_i=t^X_i(S) \oplus \left( \bigoplus_{j \in B_i} g^X_j \right) \oplus (l^X_i \cdot l_X),
\end{equation}
\begin{equation}
\label{zi_y}
z_i=t^Z_i(S) \oplus \left( \bigoplus_{j \in B_i} g^X_j \right) \oplus (l^Z_i \cdot l_Z),
\end{equation}
where $l^X_i$ and $l^Z_i$ are binary variables by which the logical operators $L_X$ and $L_Z$ are given by $L_X = \prod_i X_i^{l^X_i}$, $L_Z = \prod_i Z_i^{l^Z_i}$, respectively.
By defining $\sigma^X_j:=1-2g^X_j$, $\sigma^Z_j:=1-2g^Z_j$, $J^X_i:=(1-2l^X_i \cdot l_X)(1-2t^X_i(S))$ and $J^Z_i:=(1-2l^Z_i \cdot l_Z)(1-2t^Z_i(S))$ in Eqs.~(\ref{xi_y}) and (\ref{zi_y}), we convert the binary variables into spin variables (note that $\sigma^X_j$ and $\sigma^Z_j$ are not Pauli operators).
Then, we obtain the equations
\begin{equation}
x_i=\frac{1-J^X_i \prod_{j \in B_i} \sigma^X_j}{2}:=\frac{1-x'_i}{2},
\end{equation}
\begin{equation}
z_i=\frac{1-J^Z_i \prod_{j \in B_i} \sigma^Z_j}{2}:=\frac{1-z'_i}{2},
\end{equation}
so Eq.~(\ref{num_error_depolar}) is expressed as
\begin{align}
   &\sum_{i} \frac{1-x'_i}{2} + \sum_{i} \frac{1-z'_i}{2} - \sum_{i} \left( \frac{1-x'_i}{2} \right) \left(\frac{1-z'_i}{2}\right) \\
  &=\frac{1}{4} \sum_i \left( 3-x'_i-z'_i-x'_iz'_i\right) \\
  &=\frac{1}{4}\left( 3N- \sum_i x'_i- \sum_i z'_i- \sum_i x'_iz'_i\right),
\end{align}
where $N$ is the number of qubits.
Accordingly, the number of errors, namely Eq.~(\ref{num_error_depolar}) can be represented as an Ising Hamiltonian:
\begin{equation}
\label{hubo_depolar}
H_{l_X, l_Z} =  H_{l_X}^{X} + H_{l_Z}^{Z} + H_{l_X,l_Z}^Y,
\end{equation}
where
\begin{equation}
    H_{l_X}^{X}=-\sum_{i} J^X_i \prod_{j \in B_i} \sigma^X_j,
\end{equation}
\begin{equation}
    H_{l_Z}^{Z}=-\sum_{i} J^Z_i \prod_{j \in B_i} \sigma^Z_j,
\end{equation}
\begin{equation}
    H_{l_X,l_Z}^Y=-\sum_{i} J^X_iJ^Z_i \prod_{j \in B_i} \sigma^X_j\prod_{k \in B_i} \sigma^Z_k.
\end{equation}
Here, constant factors and terms are omitted, similar to the previous discussion.
The Hamiltonian $H_{l_X, l_Z}$ corresponds to the six-body Ising model.
By minimizing the Hamiltonian $H_{l_X, l_Z}$ in Eq.~(\ref{hubo_depolar}), it is possible to obtain an error configuration that minimizes the number of errors. 
As in the case of the bit-flip noise model, to obtain the error configuration with the minimum number of errors, we should consider each possible combination of $l_X$ and $l_Z$.
Considering the minimization of the Hamiltonian in all cases, the error configuration with the smallest Hamiltonian is the one with the minimum number of errors.
Thus, we can decode for the depolarizing noise model by solving 
\begin{equation}
\label{hubo_depolar_min}
\min _{l_X, l_Z}  \left( \mathrm{min} \; H_{l_X, l_Z} \right).
\end{equation}
Again, the decoding problem can be formulated as an energy minimization problem of an Ising Hamiltonian.
By solving Eq.~(\ref{hubo_depolar_min}) using SA, $g^X_j$ and $g^Z_j$ optimized so that the number of errors becomes minimum can be obtained.

\section{Decoding algorithm for phenomenological noise}
\label{Decoding algorithm for phenomenological noise}
\subsection{Combinatorial optimization problem for phenomenological noise}
So far, we have assumed that the error occurs only in the data qubits and that the perfect syndrome is obtained.
However, in reality the syndrome measurement is also subject to errors. 
Here, we extend the proposed decoding method to a phenomenological noise model that introduces errors into the measured syndrome phenomenologically.
The phenomenological noise model is a noise model in which bit-flip errors occur on data qubits with a probability $p$, and measured syndrome also flips with the same probability $p$.
In this situation, it is possible to achieve decoding by repeating the syndrome measurement process a number of times equal to the code distance $d$. 
For simplicity, we assume that syndrome measurement at the final round is obtained perfectly. 
Then, we obtain the error configuration that minimizes the total number of data errors and measurement errors satisfying the syndrome over all time. 
Note that the syndrome values themselves at each time step reflect the accumulation of data errors through past rounds. 
The data errors occurred at each time step can be extracted by taking 
the difference (XOR) of the syndrome values from one time step to the next.

In the following, the cumulative data error occurring in the $i$-th qubit up to time $t$ is denoted by $w^{(t)}_i \in \{0,1\}$, the measurement errors that occurred on the face $f$ by $r^{(t)}_f \in \{0,1 \}$ at time $t$, and the measured syndrome values on face $f$ by $s^{(t)}_f \in \{0,1 \}$ at time $t$. 
The measured syndrome values at time $t$ and $t-1$ can be expressed as
\begin{equation}
  \label{sft}
  s^{(t)}_f=\bigoplus_{i \in f} w^{(t)}_i \oplus r^{(t)}_f,
\end{equation}
\begin{equation}
  \label{sft_1}
  s^{(t-1)}_f=\bigoplus_{i \in f} w^{(t-1)}_i \oplus r^{(t-1)}_f.
\end{equation}
The difference $s^{(t)}_f \oplus s^{(t-1)}_f$ between $s^{(t)}_f$ and $s^{(t-1)}_f$ implies data error at time $t$.
Also, the difference between $w^{(t-1)}_i$ and $w^{(t)}_i$ corresponds to the newly occurring data error at time $t$, so we write it as $x^{(t)}_i$:
\begin{equation}
  x^{(t)}_i=w^{(t-1)}_i \oplus w^{(t)}_i.
\end{equation}
Thus, the syndrome condition that should be satisfied is 
\begin{equation}
  \label{syndphenomeno}
  \bigoplus_{i \in f} x^{(t)}_i \oplus r^{(t)}_f \oplus  r^{(t-1)}_f= s^{(t)}_f \oplus s^{(t-1)}_f, \quad \forall f.
\end{equation}
Therefore, the problem of decoding in the phenomenological noise model can be formulated as an integer programming problem with a constraint:
\begin{equation}
  \label{minphenomeno}
  \mathrm{min} \quad \sum_{i,t} x^{(t)}_i+ \sum_{f,t} r^{(t)}_f,
\end{equation}

\begin{equation}
  \label{stophenomeno}
  \mathrm{s.t.} \quad   \bigoplus_{i \in f} x^{(t)}_i \oplus r^{(t)}_f \oplus  r^{(t-1)}_f= s^{(t)}_f \oplus s^{(t-1)}_f \quad \forall f.
\end{equation}
As with the noise models we have been dealing with, this constrained combinatorial optimization problem is mapped to an unconstrained optimization problem by changing the variables.

\subsection{Decoder for phenomenological noise model}
\begin{figure}[t]
  \begin{center}
        \includegraphics[width=0.9\linewidth]{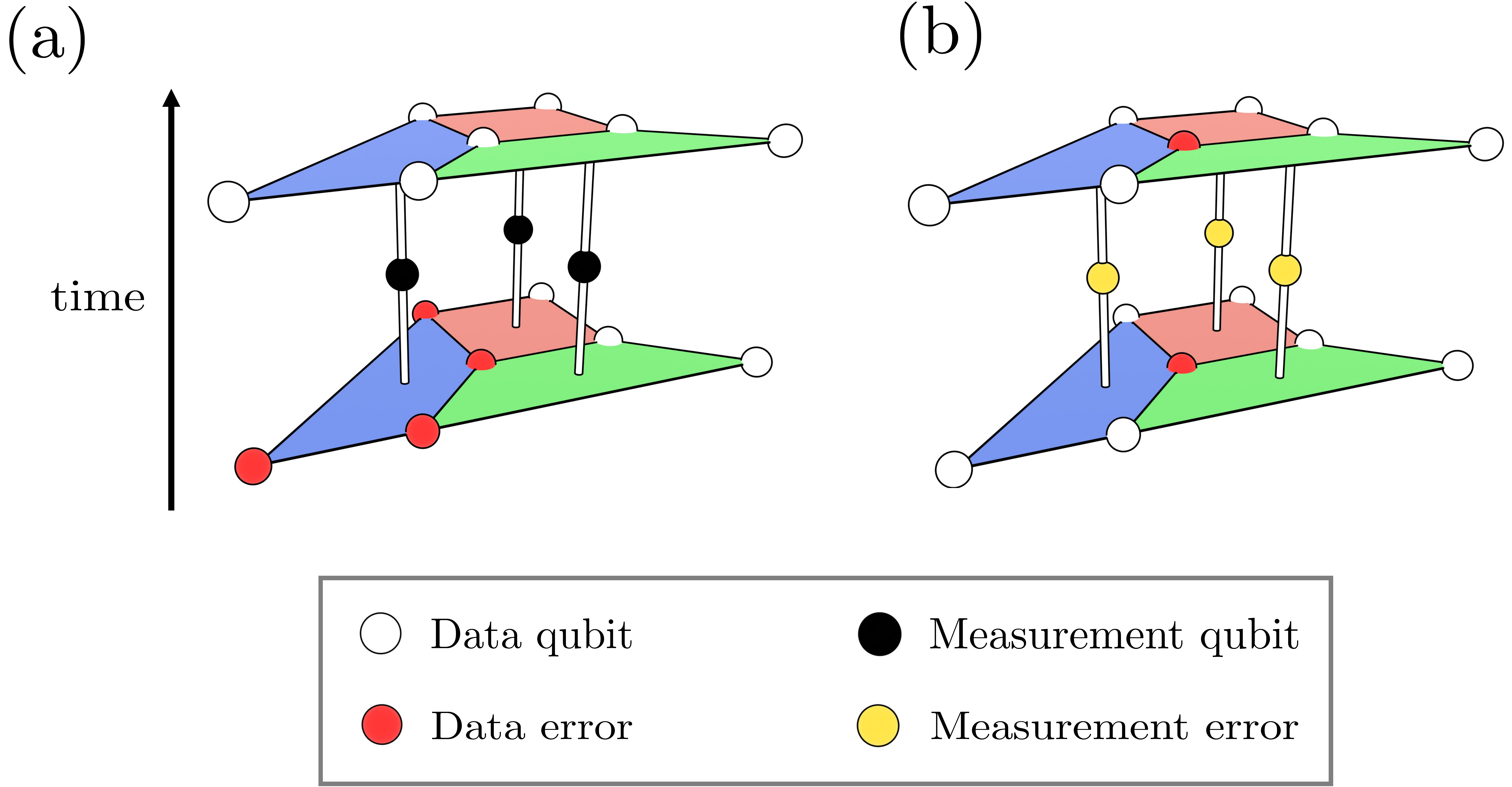}
  \end{center}
  \caption{\textbf{The space-like stabilizer generator and the time-like stabilizer generator.} (a) The space-like stabilizer generator. The four data errors indicate the action of space-like stabilizer generators defined on a blue face. (b) The time-like stabilizer generator. A data error in a qubit occurs followed by measurement errors in all faces containing that qubit, and then a data error occurs in the same qubit at the next time.}
  \label{fig:space-like_timelike}
\end{figure}

Even in situations where temporal errors are considered, data errors represented as binary variables can be expressed in a similar form as Eq.~({\ref{xi}}).
We can decompose temporal data error vector $\bm{x}_i$:
\begin{equation}
    \label{x_vector}
    \bm{x}_i=\bar{\bm{t}}_i(S) \oplus \bm{g}_i \oplus \bm{l}_i,
\end{equation}
where each $t$-th element of the vectors $\bm{x}_i$, $\bar{\bm{t}}_i(S)$, $\bm{g}_i$, 
and $\bm{l}_i$, corresponds to variables at time $t$, e.g. $x_i^{(t)}$. 
$S$ denotes the syndrome for all time steps.
Regarding the pure error, we should consider operations in each time slice in a similar way to the bit-flip noise model.
We write a bit that represents the action of a pure error on the $i$-th qubit at time $t$ as $\bar{t}^{(t)}_i(S)$.
With regard to stabilizer operators, the previous approach alone does not work in this case.
In the presence of measurement errors, 
we have to consider not only stabilizer generators defined on each time slice but also 
spatio-temporal error events to cover 
all transformations that preserve the space-time syndrome values.
Here, we call the stabilizer generators acting in the same time slice as ``space-like stabilizer generators", 
and the stabilizer generators that act along the time axis as ``time-like stabilizer generators", 
as described in Fig.~\ref{fig:space-like_timelike}.

Now, in order to prove the decomposition as shown in Eq.~(\ref{x_vector}), we provide a proof that any error event including data errors and measurement errors is expressed as a product of a pure error, a space-like stabilizer operator, a time-like stabilizer operator, and a logical operator. 
To prove this, it is sufficient to show that any error event including data error and measurement error that makes the left side of Eq.~(\ref{syndphenomeno}) equal to zero can always expressed as a product of a space-like stabilizer operator, a time-like stabilizer operator, and a logical operator. 
This is because the quantum state after application of pure errors at each time slice has trivial space-time syndrome values.
Firstly, by the definition of time-like stabilizer generators which will be explained in more detail in the following paragraph, any error events including arbitrary patterns of $r^{(t)}_f$ for all $f$ and $t$ that make the left side of Eq.~(\ref{syndphenomeno}) equal to zero can be expressed using time-like stabilizer generators. 
Then, for each pattern of $r^{(t)}_f$ for all $f$ and $t$, every patterns of $x^{(t)}_i$ for all $i$ and $t$ that make the left side of Eq.~(\ref{syndphenomeno}) equal to zero can be expressed using space-like stabilizer operators and logical operators, similar to the discussion of Eq.~(\ref{tgl}). 
Therefore, it is proven that any error event including data errors and measurement errors is expressed as a product of a pure error, a space-like stabilizer operator, a time-like stabilizer operator, and a logical operator, which in turn show that the temporal data error vector $\bm{x}_i$ can be decomposed as Eq.~(\ref{x_vector}).

Time-like stabilizer generators are operators that generate a data error at time $t$ in a qubit followed by measurement errors in all faces containing that qubit, and then a data error occurs in the same qubit at time $t+1$.
There are $n$ time-like stabilizer generators between each pair of time steps, so there are $n(d-1)$ of them in total for all time steps.
With this simple definition, the order of the interaction of the Hamiltonians is minimized and also it becomes easy to construct the Hamiltonians.
Then, by utilizing both time-like and space-like stabilizer generators, the action of the stabilizer operator to the $i$-th qubit at time $t$ can be represented as
\begin{equation}
  \label{stbphenomeno}
g^{(t)}_i=\left(\bigoplus_{j \in B_i} {g'}^{(t)}_j \right) \oplus \bar{g}^{(t)}_i  \oplus \bar{g}^{(t-1)}_i,
\end{equation}
where ${g'}^{(t)}_j$ is a binary variable representing whether or not the space-like stabilizer generator at time $t$ is included, and 
$\bar{g}^{(t)}_i$ is a binary variable representing whether or not the time-like stabilizer generator between time $t$ and $t + 1$ is included.
$B_i$ denotes the set of indices of space-like stabilizer generators acting on the $i$-th qubit, and this set does not depend on the time $t$.
Note that the logical operator can be defined at an arbitrary one time slice, since 
its time can be changed by using time-like stabilizers. 
Therefore, the action of the logical operator to the $i$-th qubit at time $t$ can be represented as
\begin{equation}
  \label{lgphenomeno}
  l^{(t)}_i=l'^{(t)}_i \cdot l,
\end{equation}
where $l$ is a binary variable representing whether or not the logical operator at a certain time is included, $l'^{(t)}_i$ is a bit that indicates on which qubits and at what time the logical operator acts.
Here, $l$ does not need to depend on time $t$.
Then, the data error $x^{(t)}_i$ at time $t$ can be written as
\begin{align}
  \label{xphenomeno}
  x^{(t)}_i&=\bar{t}^{(t)}_i (S) \oplus g^{(t)}_i \oplus l^{(t)}_i \\
  &=\bar{t}^{(t)}_i (S) \oplus \left(\bigoplus_{j \in B_i} {g'}^{(t)}_j \right) \oplus \bar{g}^{(t)}_i  \oplus \bar{g}^{(t-1)}_i \oplus (l'^{(t)}_i \cdot l).
\end{align}
Also, measurement errors can be represented using time-like stabilizer generators. The measurement error $r^{(t)}_f$ is represented as
\begin{equation}
  \label{rphenomeno}
   r^{(t)}_f=\bigoplus_{i \in f} \bar{g}^{(t)}_i.
\end{equation}
By defining ${\sigma'}^{(t)}_j:=1-2{g'}^{(t)}_j,\ \bar{\sigma}^{(t)}_i:=1-2\bar{g}^{(t)}_i$,
and $J^{(t)}_i:=(1-2l'^{(t)}_i \cdot l)(1-2\bar{t}^{(t)}_i(S))$ to transform binary variables into Ising spin variables, the total number of errors is represented as
\begin{equation}
\label{hubo_phenomeno}
H_{l, \rm pheno} =  -\sum_{i,t} J^{(t)}_i \prod_{j \in B_i} {\sigma'}^{(t)}_j \bar{\sigma}^{(t)}_i \bar{\sigma}^{(t-1)}_i -\sum_{f,t} \prod_{i \in f}  \bar{\sigma}^{(t)}_i.
\end{equation}
By minimizing Eq.~(\ref{hubo_phenomeno}), it is possible to obtain an error configuration that minimizes the number of errors. 
Similar to the previous discussion, to obtain the error configuration with the minimum number of errors, we should consider each case, whether $l$ is 1 or 0.
Considering the minimization of the Hamiltonian in the two cases, the error configuration with the smaller Hamiltonian is the one with the minimum number of errors.
Thus, we can decode for the phenomenological noise model by solving 
\begin{equation}
  \label{hubo_phenomeno_min}
\min _{l}  \left( \mathrm{min} \; H_{l, \rm pheno} \right).
\end{equation}
Again, the decoding problem can be formulated as an energy minimization problem of an Ising Hamiltonian. The Hamiltonian $H_{l, \rm pheno}$ corresponds to the eight-body Ising model. By solving Eq.~(\ref{hubo_phenomeno_min}) with SA,
${g'}^{(t)}_j$ and $\bar{g}^{(t)}_j$ optimized so that the number of errors becomes minimum can be obtained.
This decoder can also be applied for the circuit-level noise model, with the ability to adjust the weights of data errors and measurement errors based on the probability distribution after error propagation within the circuit. By changing the coefficients of each term in the Ising Hamiltonian, it is possible to adjust the weights because they correspond to the weights of each data error and measurement error.

\section{Numerical experiments}
\label{Numerical experiments}
\subsection{Settings}
In order to evaluate the performance of the proposed decoding methods,
we perform Monte Carlo simulations to estimate the logical error rates. 
In the Monte Carlo simulations, errors were generated on the data qubits, and syndrome measurement was performed
with and without errors depending on the noise models.
Then, the Ising Hamiltonian was constructed from the observed syndrome for each of the noise models
and was solved by SA.
Specifically,
we employed the open-source library OpenJij in Python \cite{openjij}. 
The solution is used to perform a recovery operation and to see whether or not error correction fails.
These Monte Carlo simulations are repeated $10^5$ times to estimate the logical error rates accurately. 
Moreover, in order to validate the accuracy of our method, 
we performed the same task to estimate the logical error rates 
by solving the constrained optimization problem exactly by using an integer programming solver, CPLEX. 

Let us describe the details of the schedule employed in SA.
The initial spin configuration is randomly determined. 
Also, we employed annealing parameters, 
such as how to schedule temperatures, how many repetitions of spin updates are taken at each temperature, and how many iterations of the overall process are performed to obtain the best solution of them. In SA, there is a trade-off relationship between accuracy and computation time.
We used the annealing parameters realizing the shortest possible computation time while keeping the accuracy reasonably good compared to that obtained by CPLEX. 
Specifically, we adopted the annealing parameters as follows. 
The initial and final values of the inverse temperature are $\beta_{\mathrm{min}} = \log2 / \Delta E_{\mathrm{min}}$ and $\beta_{\mathrm{max}} = \log100 / \Delta E_{\mathrm{max}}$, respectively,
where $\Delta E_{\mathrm{min}}$ is a rough lower bound of the energy gap of the Hamiltonians and $\Delta E_{\mathrm{max}}$ is an upper bound of the energy gap of the Hamiltonians. These values are determined by the coefficients of each term in the Hamiltonians and are the values commonly adopted in OpenJij, respectively.
The number of Monte Carlo steps at each inverse temperature is taken to be one, 
where one Monte Carlo step means to update all spins once. 
We used the exponential cooling schedule to compute the temperature values at each temperature cycle.
The number of temperature cycles and the number of iterations are changed
depending on the noise models and code distances to obtain high-speed decoding with high accuracy as much as possible.
By thoroughly examining almost all combinations of parameters at each noise model and code distance, we found such annealing parameters.
We carried out the decoding using the same annealing parameters for all physical error rates for simplicity.
Specifically, in order to achieve high accuracy, 
we adopted the annealing parameters that can accurately decode with the physical error rate close to the threshold, where the optimization problem becomes most difficult. 
The details of how to determine the parameters we used are shown in 
Figs.~\ref{fig:parameter_bitflip}-\ref{fig:parameter_phenomeno} in Appendix~\ref{appendix}.

\subsection{Bit-flip noise model}
\begin{figure}[t]
  \centering
       \includegraphics[width=0.9\linewidth]{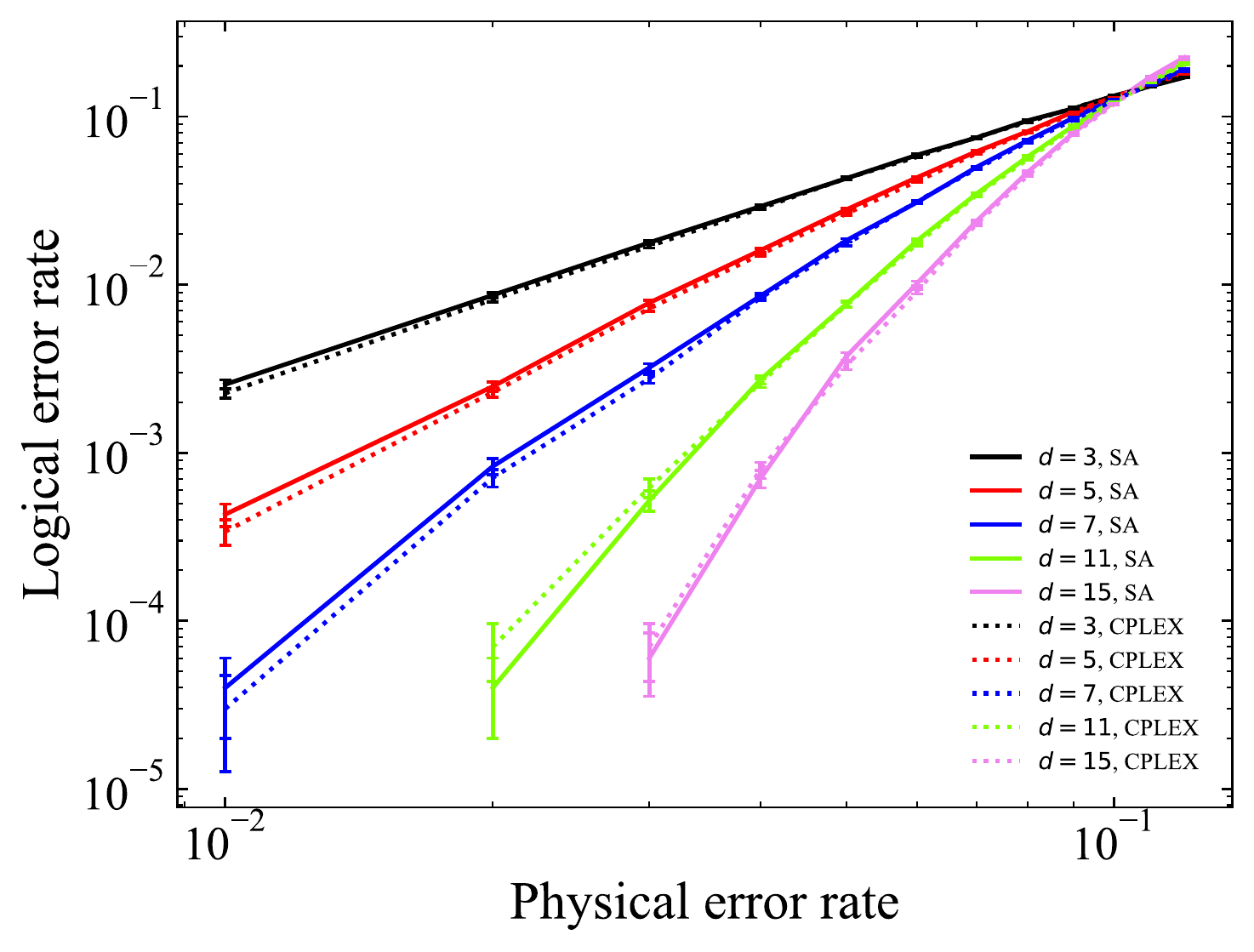}
  \caption{\textbf{Numerical simulations of logical error rate for the bit-flip noise model for various code distances.} The solid line data represents our data and the dotted line data represents the data from CPLEX. The threshold obtained by our method is 10.36(5)\%}
  \label{fig:error_bitflip}
\end{figure}

The resultant logical error rates for $d=3,5,7,11,15$ under the bit-flip noise model are plotted 
as functions of a physical error rate $p$ in Fig.~\ref{fig:error_bitflip}.
The annealing parameters of SA are listed in Table~\ref{tab:sa_parameter_bitflip}.
It is known that around the threshold and for sufficiently large code distance $d$, the logical error rate $p_\mathrm{L}$ should scale as
\begin{equation}
\label{threshold_scale}
    p_\mathrm{L}=(p-p_{\mathrm{th}})d^{1/\nu_0}.
\end{equation}
Considering finite-size effects, we fit our data to the form
\begin{equation}
\label{threshold_scale_revised}
    p_{\mathrm{L}}=A+B\left( p-p_{\mathrm{th}} \right)d^{1/{\nu_0}},
\end{equation}
instead of Eq.~(\ref{threshold_scale}).
As a result, we obtained
\begin{equation}
  p_{\mathrm{th}}=0.1036 \pm 0.0005,
\end{equation}
\begin{equation}
  \nu_0=1.2 \pm 0.2,
\end{equation}which is close to the threshold value of 10.56(1)\% found by the exact integer program decoder based on the concept of minimum distance decoding in Ref.~\cite{landahl}.
 Also, our threshold is higher than any other decoder in prior studies, for example, 8.7\% for the renormalization group decoder \cite{renormalize}, 10.2\% for the restriction decoder with MWPM \cite{unionfind}, and 9.8\% (analytical estimate) for the restriction decoder with union-find \cite{unionfind}.
Although our threshold is lower than the optimal threshold of 10.925(5)\% \cite{x_threshold}, it is not a surprising result. 
While color codes are degenerate codes, the proposed method is based on the concept of minimum distance decoding and does not consider the degeneracy of error configurations, contributing to the slightly lower threshold.

Regarding the critical exponent, it belongs to the same universality class as the Ising critical exponent, which characterizes the behavior of the Ising model near its critical point \cite{scaling2}.
Although our critical exponent $\nu_0$ is nearly consistent with the value of $\nu_0$=1.463(6) found for the surface code \cite{scaling2}, due to the large impact of statistical errors, it is essentially unclear whether these values are consistent or not. 
\begin{table}[t]
    \caption{The parameters of SA used for decoding the bit-flip noise model.\label{tab:sa_parameter_bitflip}}
    \begin{ruledtabular}
        \begin{tabular}{ccc}
        $d$ & \# of temperature cycles&\# of iterations\\
        3&30&5\\
        5&70&5\\
        7&50&10\\
        11&100&20\\
        15&450&20
        \end{tabular}
    \end{ruledtabular}
\end{table}
\begin{figure}[t]
  \begin{center}
    \includegraphics[width=0.95\linewidth]{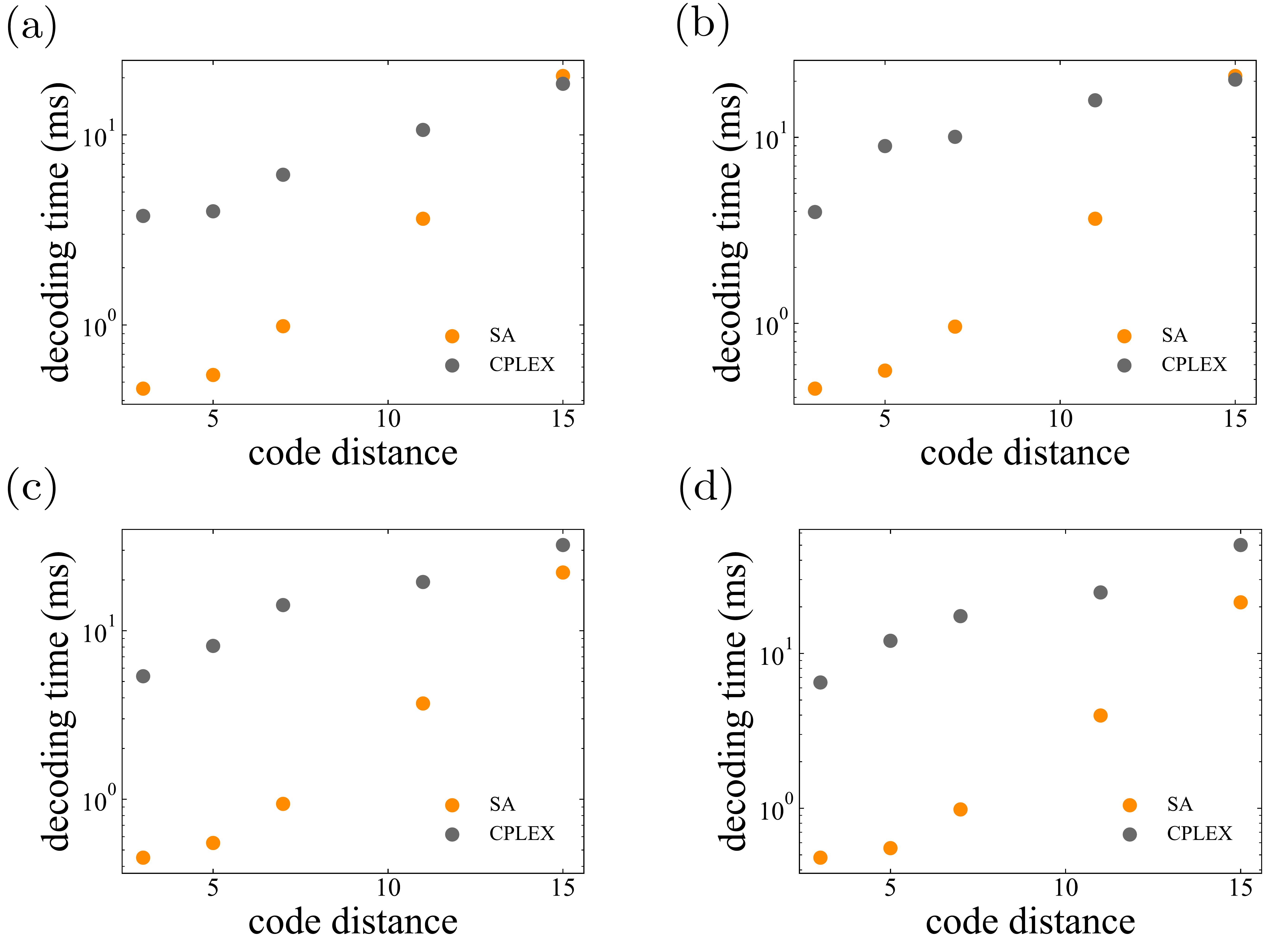}
  \end{center}
  \caption{\textbf{Decoding time for the bit-flip noise model.} (a) $p=0.01$. (b) $p=0.04$. (c) $p=0.07$. (d) $p=0.10$.}
  \label{fig:bitflip_time}
\end{figure}
Compared to the logical error rates obtained by CPLEX, 
the proposed decoder provides almost the same accuracy within statistical errors,
which clearly shows that the proposed decoder with SA works well achieving minimum distance decoding.

To compare the decoding time of SA and CPLEX,
we show the decoding time for code distances $d=3,5,7,11,15$ with the physical error rates $p=0.01$ (a), $p=0.04$ (b), $p=0.07$ (c), and $p=0.10$ (d)  in Fig.~\ref{fig:bitflip_time}.
To make a fair comparison, we compared the computation time using a single core of a CPU for both cases,
while our method can be simply parallelized when multiple cores are available.
The proposed method with SA succeeded to decode faster than CPLEX when the physical error rate is around the threshold or when $d$ is small.
The reason why the proposed method may be slower than CPLEX in cases where the physical error rate is low is that the SA parameters are optimized around the threshold. 
 If we optimize the annealing parameters for each physical error rate, we can decode faster even in cases of low physical error probability, while this is out of scope in this work.
Also, we can decode faster by taking advantage of the parallel computation of SA. 
SA can be parallelized with 100\% efficiency for iterations. 
In addition, each iteration can be parallelized and accelerated using GPUs \cite{sa_gpu}.

\subsection{Depolarizing noise model}
\begin{figure}[t]
  \centering
    \includegraphics[width=0.9\linewidth]{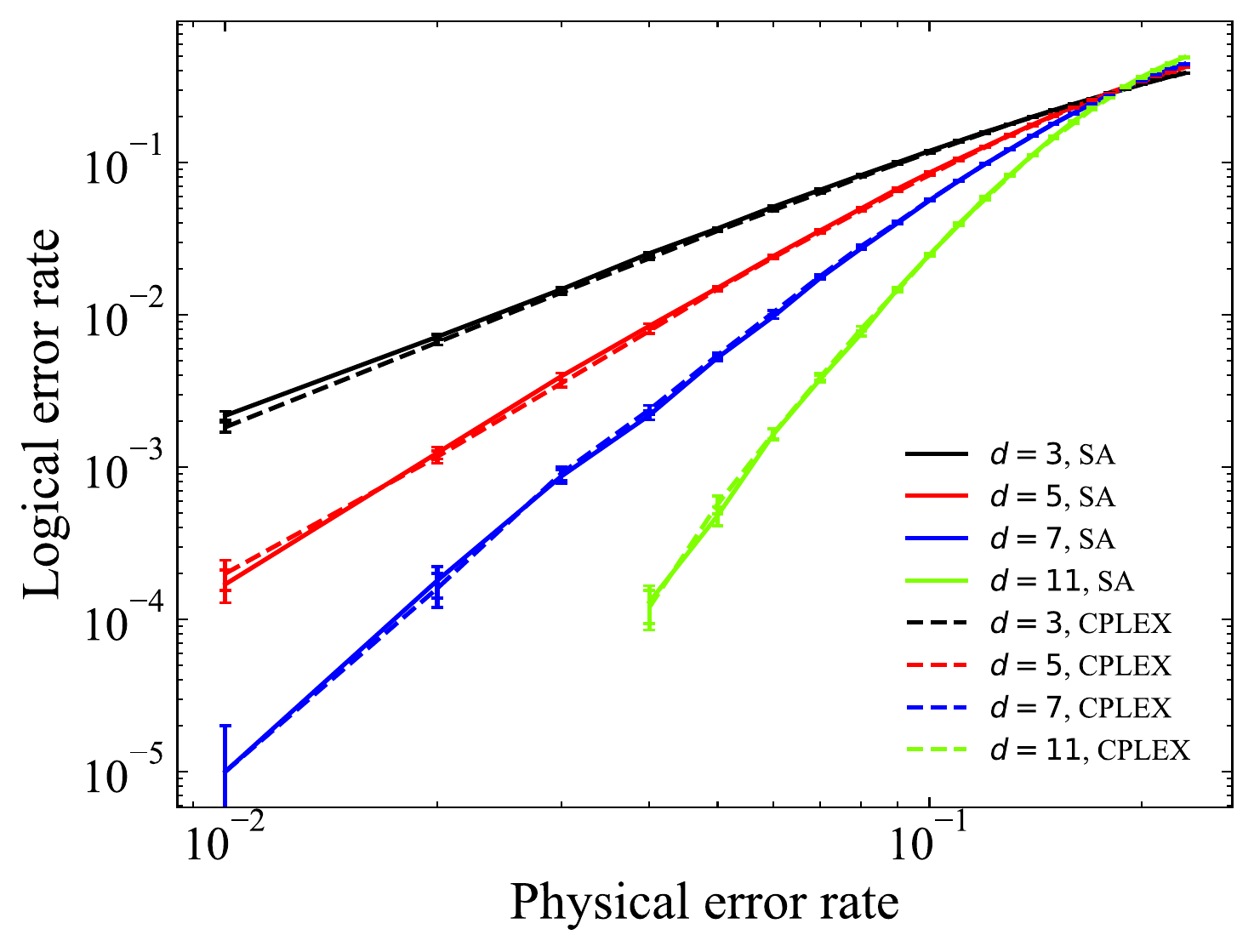}
  \caption{\textbf{Numerical simulations of logical error rate for the depolarizing noise model for various code distances.} The threshold obtained by our method is 18.47(5)\%}
  \label{fig:error_depolar}
\end{figure}

Next, we show the logical error rates for $d=3,5,7,11$ as functions of a physical error rate $p$ 
under the depolarizing noise model in Fig.~\ref{fig:error_depolar}.
The parameters of SA used in this process are listed in Table~\ref{tab:sa_parameter_depolar}.
Again, we fit our data to the formula Eq.~(\ref{threshold_scale_revised}). We found
\begin{equation}
  p_{\mathrm{th}}=0.1847 \pm 0.0005,
\end{equation}
\begin{equation}
  {\nu_0}=1.3 \pm 0.2,
\end{equation}which corresponds closely to the threshold value of 18.6(3)\% that we estimated by the integer program decoder.
Our threshold is higher than the threshold value of 17.5\% found by the neural-network decoder \cite{neural}. 
For the same reasons as in the case of the bit-flip noise model, our threshold is less than the value of 18.9\% for optimal decoding \cite{xyz_threshold}.
 Moreover, similar to the previous case, 
the logical error rates of the proposed decoder are almost the same as those obtained by CPLEX.
\begin{table}[t]
    \caption{The parameters of SA used for decoding the depolarizing noise model.\label{tab:sa_parameter_depolar}}
    \begin{ruledtabular}
        \begin{tabular}{ccc}
        $d$ & \# of temperature cycles&\# of iterations\\
        3&150&5\\
        5&100&10\\
        7&200&20\\
        11&1200&30
        \end{tabular}
    \end{ruledtabular}
\end{table}

\begin{figure}[t]
  \begin{center}
    \includegraphics[width=0.95\linewidth]{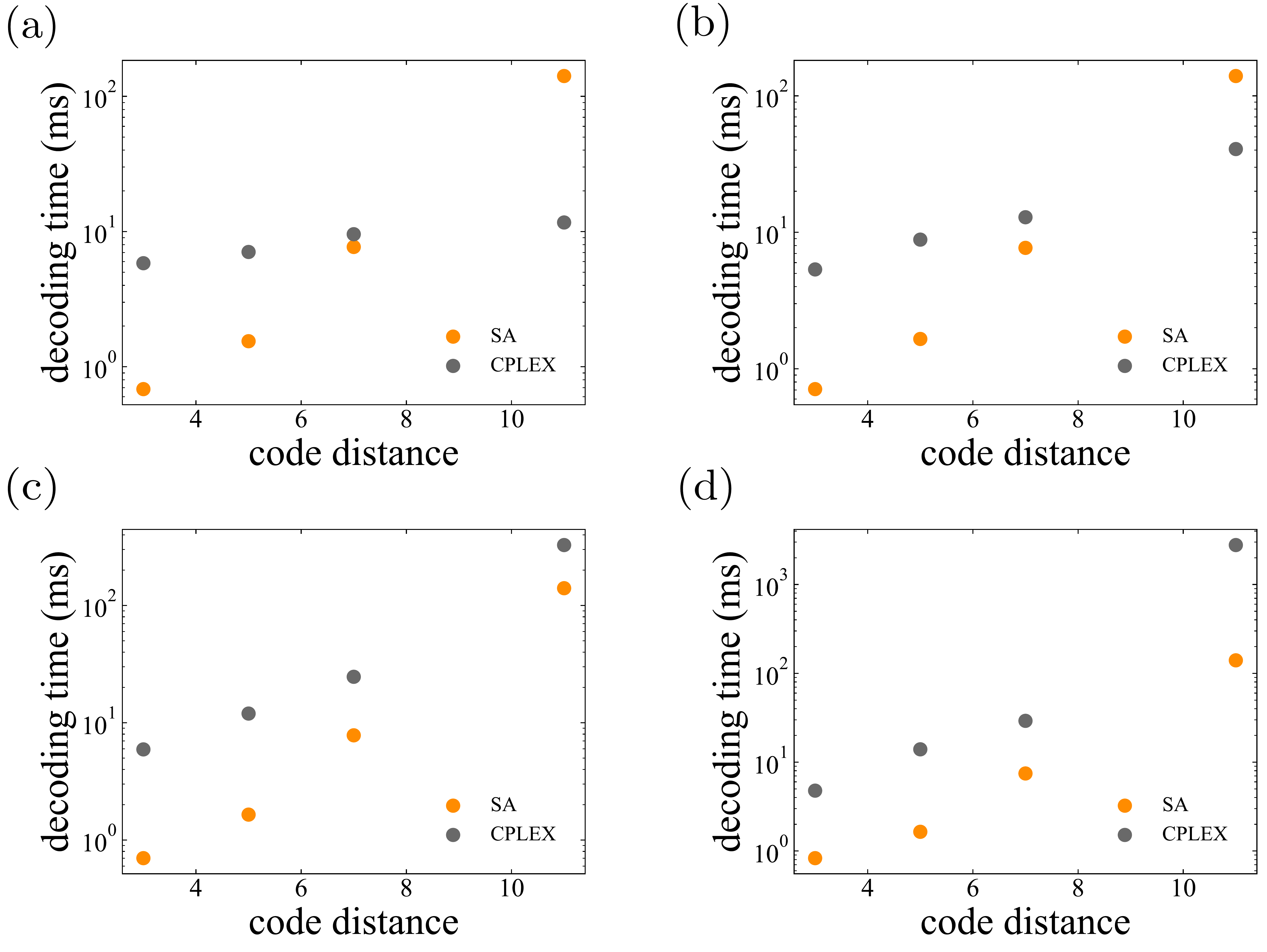}
  \end{center}
  \caption{\textbf{Decoding time for the depolarizing noise model.} (a) $p=0.01$. (b) $p=0.07$. (c) $p=0.13$. (d) $p=0.18$.}
  \label{fig:depolar_time}
\end{figure}

We demonstrate the decoding time measured using a single core of the CPU for code distances $d=3,5,7,11$ with the physical error rates $p=0.01$ (a), $p=0.07$ (b), $p=0.13$ (c), and $p=0.18$ (d) in Fig.~\ref{fig:depolar_time}. 
Similar to the case of the bit-flip noise model, our method achieved shorter computation time compared to CPLEX when either the physical error rate is around the threshold, or when $d$ is not large.

\subsection{Phenomenological noise model}
\begin{figure}[t]
  \centering
    \includegraphics[width=0.9\linewidth]{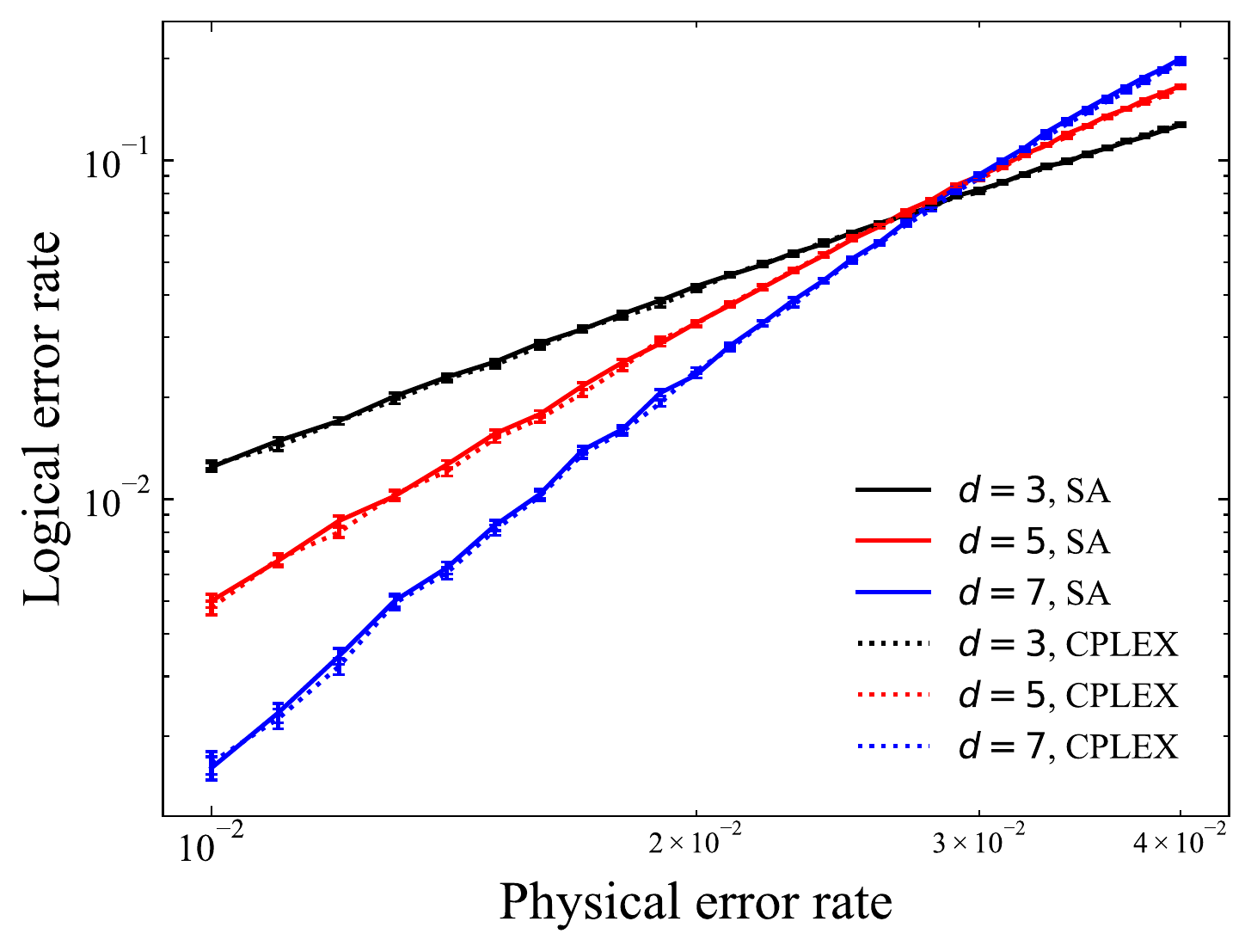}
  \caption{\textbf{Numerical simulations of logical error rate for the phenomenological noise model for various code distances.} The threshold obtained by our method is 2.90(4)\%}
  \label{fig:error_phenomeno}
\end{figure}

Finally, we show the logical error rates for $d=3,5,7$
as a functions of a physical error rate $p$ under the phenomenological noise model in Fig.~\ref{fig:error_phenomeno}. In this noise model, bit-flip errors occur on data qubits with a probability $p$, and measured syndrome also flips with the same probability $p$.
The parameters of SA used in this process are listed in Table~\ref{tab:sa_parameter_phenomeno}.
\begin{figure}[t]
  \begin{center}
    \includegraphics[width=0.95\linewidth]{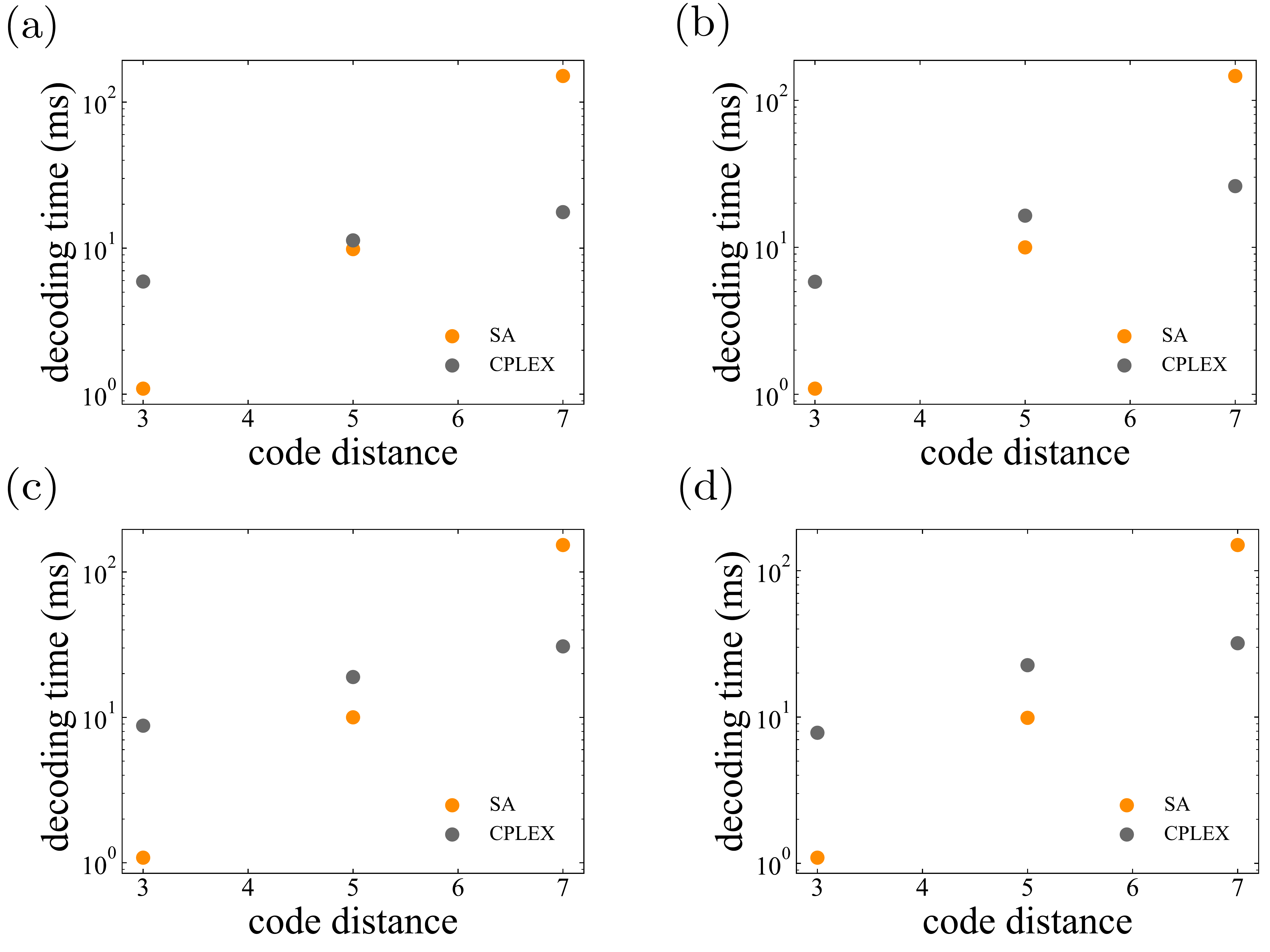}
  \end{center}
  \caption{\textbf{Decoding time for the phenomenological noise model.} (a) $p=0.010$. (b) $p=0.018$. (c) $p=0.024$. (d) $p=0.028$.}
  \label{fig:phenomeno_time}
\end{figure}
\begin{table}[t]
    \caption{The parameters of SA used for decoding the phenomenological noise model.\label{tab:sa_parameter_phenomeno}}
    \begin{ruledtabular}
        \begin{tabular}{ccc}
        $d$ & \# of temperature cycles&\# of iterations\\
        3&150&5\\
        5&100&20\\
        7&300&40
        \end{tabular}
    \end{ruledtabular}
\end{table}
Again, we fit our data to the form Eq.~(\ref{threshold_scale_revised}). We found
\begin{equation}
  p_{\mathrm{th}}=0.0290 \pm 0.0004,
\end{equation}
\begin{equation}
  {\nu_0}=1.3 \pm 0.2,
\end{equation}which is in good agreement with the threshold values of 3.05(4)\% estimated by the integer program decoder in Ref.~\cite{landahl}. 
Compared to other previous studies, our threshold is higher than the value of 2.08\% estimated by graph matching decoder \cite{graphmatching}.
The critical exponent also closely agrees with the value of $\nu_0=1.5(2)$ previously estimated by the integer program decoder \cite{landahl}.
As with the bit-flip and depolarizing noise model, the proposed decoder shows logical error rates that are comparable to those obtained by CPLEX.

The decoding time for code distances $d=3,5,7$ with the physical error rates $p=0.010$ (a), $p=0.018$ (b), $p=0.024$ (c), and $p=0.028$ (d) is shown in Fig.~\ref{fig:phenomeno_time}. 
In this noise model, we succeeded to decode faster than CPLEX when $d$ is small for almost all physical error rates. 
While our method is slower than CPLEX with large $d$, OpenJij can be further optimized in solving the energy minimization problem of the multi-body Ising Hamiltonian. 
In the case of the phenomenological noise model, the Hamiltonian consists of eight bodies. It is believed that we can decode faster by constructing a SA solver that is optimized for the multi-body Ising Hamiltonians instead of using OpenJij.
Also, similar to the previous discussion, we can decode faster by tuning the parameters depending on the physical error rate or performing parallel computation.

\section{Conclusion}
\label{conclusion}
In this paper, we proposed an Ising model formulation for highly accurate color codes decoding, which is solved by SA.
The decoding results show that our method can achieve almost the same accuracy as the integer program decoder using CPLEX for all noise models we have employed here if the annealing schedules are appropriately chosen. 
The decoding time is smaller than CPLEX when the physical error rate is around the threshold or when $d$ is small for the bit-flip and depolarizing noise model. Also, for the phenomenological noise model, we succeeded to decode faster than CPLEX when $d$ is small for almost all physical error rates. 

In this work, the number of temperature cycles and the number of iterations in the SA are only optimized. 
However, the other parameters such as the initial and target inverse temperature, the number of Monte Carlo steps at each inverse temperature, etc., are not optimized but set to be the default values in OpenJij.
By setting the target inverse temperature to the Nishimori temperature \cite{nishimori1981internal,fujii2015quantum}, the proposed decoder leads to a performance that is closer to the optimal one, taking into consideration the degeneracy.
The proper setting of such annealing parameters has been a long-standing research topic, and several methods have been proposed \cite{sa1,sa2}, but no optimal method has been established. 
Therefore, we can decode more accurately and faster by further optimizing the annealing parameters. 
In this study, we did not optimize the SA solver itself for our purpose, but used the versatile open-source software OpenJij.
However, we can further reduce the decoding time by developing an SA solver specific to the Ising model associated with the color code decoding problem and by parallelizing it with multi-core CPUs and GPUs.
Additionally, instead of selecting parameters that achieve high performance, we can use parameters that result in lower performance but shorter decoding time. This is because our decoder using SA is in a trade-off relationship between performance and decoding time.

Here we should note that it is a rare result that we can solve combinatorial optimization problems through a heuristic approach with a smaller time compared to exact solvers \cite{ohzeki2019control}.
The reason is that in general, when solving combinatorial optimization problems using SA, the problem needs to be embedded into an Ising form, which leads to a loss. 
Nevertheless, our study achieved favorable results with SA, which can be attributed to the compatibility between SA and the error correction problem. 
In our method, the error correction problem is formulated as an Ising problem, and SA can be executed without the loss of problem embedding, resulting in a good performance.

Constructing high-performance decoding methods for the color codes presented here 
has been a challenging task, unlike the surface codes. 
However, our proposed decoding method enables highly accurate decoding to be achieved in a shorter amount of time compared to other decoders achieving comparable performance.
Although color codes are still suffering from low thresholds in the circuit-level noise model, our method simplifies threshold estimation and makes it easier to improve the performance through trial and error in the architecture design.
Recently, QEC on small color codes has been experimentally implemented \cite{bluvstein2023logical}. In such near-term QEC experiments, there is a need to demonstrate that decoders can, in principle, handle correlated errors that often arise when applying e.g., transversal entangling gates. It is reasonable to spend a relatively large amount of time on decoding for this purpose, particularly around the threshold. In this situation, our decoder especially has advantages, as it allows adjusting performance and decoding time within a trade-off by tuning parameters.
Furthermore, the decoding method proposed in this paper can be readily applied to any stabilizer code
and hence has a great potential to advance the field of QEC, including quantum low-density parity check codes \cite{breuckmann2021quantum}, where the decoding problem is highly non-trivial. 
Further studies can explore the scalability of the method and its effectiveness in real-world scenarios, with the aim of improving the performance of quantum computing systems.

\begin{acknowledgments}
The authors would like to thank K. Suzuki, K. Nishimura, and Y. Yamashiro for valuable discussion and implementation of OpenJij. 
This work is supported by
MEXT Quantum Leap Flagship Program (MEXT Q-LEAP)
Grant No. JPMXS0118067394 and JPMXS0120319794, JST
COI-NEXT Grant No. JPMJPF2014, and JST Moonshot
R\&D Grant No. JPMJMS2061.
\end{acknowledgments}

\appendix
\section{Relationship between annealing parameters and logical error rates.}
\label{appendix}
\begin{figure}[b]
  \begin{center}
    \includegraphics[width=\linewidth]{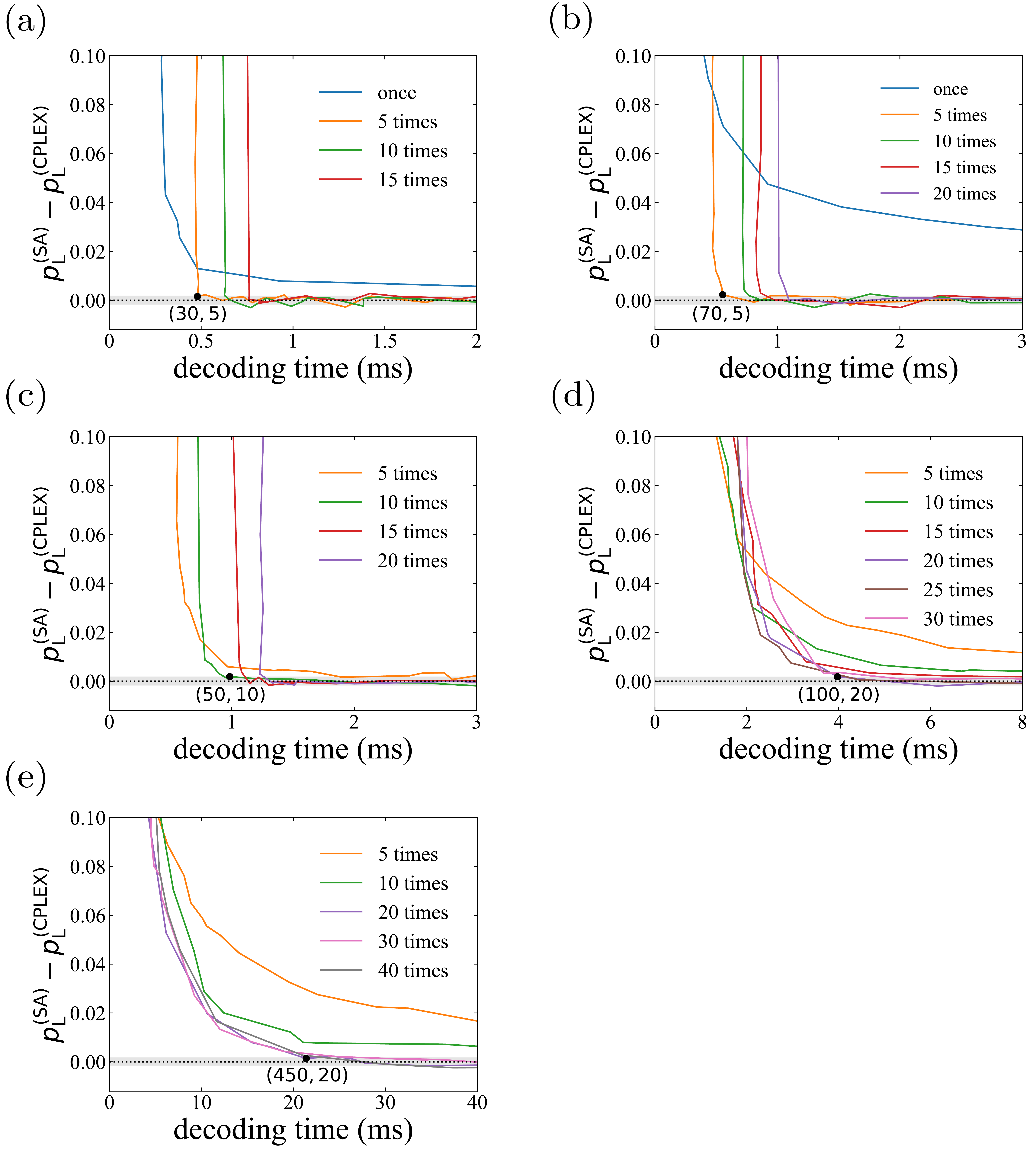}
  \end{center}
  \caption{\textbf{Dependence of $p^{(\mathrm{SA})}_{\mathrm{L}}-p^{(\mathrm{CPLEX})}_{\mathrm{L}}$ on annealing parameters in the bit-flip noise model.} (a) $d=3$. (b) $d=5$. (c) $d=7$. (d) $d=11$. (e) $d=15$. The physical error rate assumed in this figure is 10.0\%.}
  \label{fig:parameter_bitflip}
\end{figure}
Here we explain about how to determine the number of temperature cycles and the number of iterations to achieve high accuracy with minimum computational overhead. 
In minimum distance decoding, the logical error rates achieved by CPLEX are theoretically the lowest, 
so the accuracy of our decoding can be evaluated by examining how close the logical error rates achieved by SA are to the logical error rates achieved by CPLEX. 
If the physical error rate $p$ is low, as described in Ref.~\cite{scaling1}, the logical error rate achieved by SA should scale as
\begin{equation}
  p^{(\mathrm{SA})}_{\mathrm{L}}=c^{(\mathrm{SA})}\left( \frac{p}{p^{(\mathrm{SA})}_{\mathrm{th}}} \right)^{\frac{d+1}{2}},
\end{equation}
and the logical error rate achieved by CPLEX should also scale as
\begin{equation}
  p^{(\mathrm{CPLEX})}_{\mathrm{L}}=c^{(\mathrm{CPLEX})}\left( \frac{p}{p^{(\mathrm{CPLEX})}_{\mathrm{th}}} \right)^{\frac{d+1}{2}},
\end{equation}
where $c^{(\mathrm{SA})}$, $c^{(\mathrm{CPLEX})}$ denotes constants, and $p^{(\mathrm{SA})}_{\mathrm{th}}$, $p^{(\mathrm{CPLEX})}_{\mathrm{th}}$ represent the threshold for SA and CPLEX, respectively. 
Thus, we can use the expression 
\begin{equation}
\small
\label{ratio}
  \left(\frac{p^{(\mathrm{SA})}_{\mathrm{L}}}{p^{(\mathrm{CPLEX})}_{\mathrm{L}}} \right)^{\frac{2}{d+1}}= \left( \frac{c^{(\mathrm{SA})}}{c^{(\mathrm{CPLEX})}}  \right)^{\frac{2}{d+1}} \left( \frac{p^{(\mathrm{CPLEX})}_{\mathrm{th}}}{p^{(\mathrm{SA})}_{\mathrm{th}}} \right)
\end{equation}
as an indicator to evaluate the decoding accuracy of SA, where the reason for the $2/(d+1)$ power is to reduce the $d$-dependence.
The closer Eq.~(\ref{ratio}) is to 1, the higher the decoding accuracy of SA.
On the other hand, around the threshold where the error probability is high, the difference between $p^{(\mathrm{SA})}_{\mathrm{L}}$ and $p^{(\mathrm{CPLEX})}_{\mathrm{L}}$ may serve as an indicator for evaluating the decoding accuracy of SA.
The closer the difference between $p^{(\mathrm{SA})}_{\mathrm{L}}$ and $p^{(\mathrm{CPLEX})}_{\mathrm{L}}$ is to 0, the higher the accuracy of the decoding by SA.

\begin{figure}[htb]
  \begin{center}
    \includegraphics[width=\linewidth]{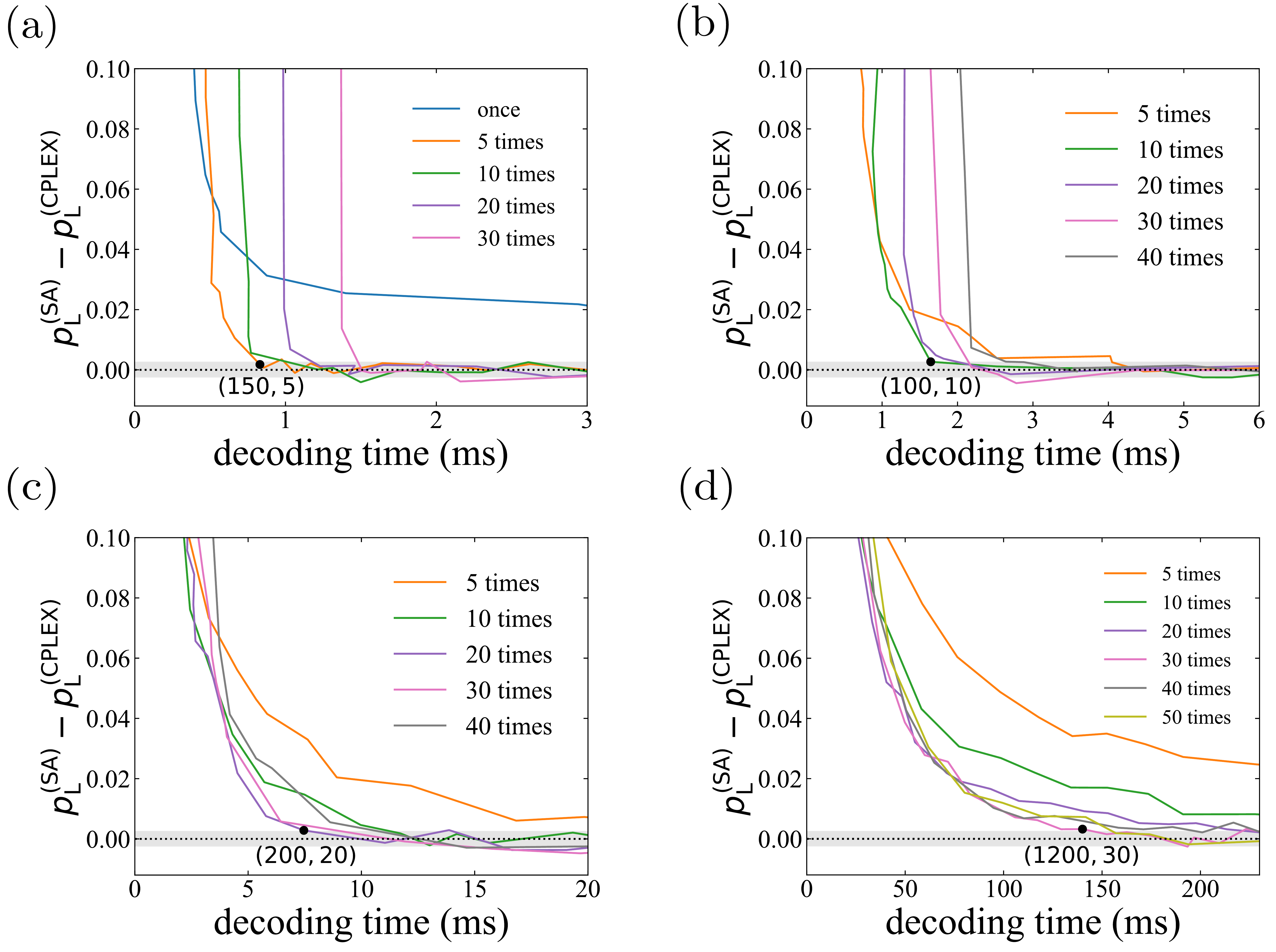}
  \end{center}
  \caption{\textbf{Dependence of $p^{(\mathrm{SA})}_{\mathrm{L}}-p^{(\mathrm{CPLEX})}_{\mathrm{L}}$ on annealing parameters in the depolarizing noise model.} (a) $d=3$. (b) $d=5$. (c) $d=7$. (d) $d=11$. The physical error rate assumed in this figure is 18.0\%.}
  \label{fig:parameter_depolar}
\end{figure}
\begin{figure}[tb]
  \begin{center}
    \includegraphics[width=\linewidth]{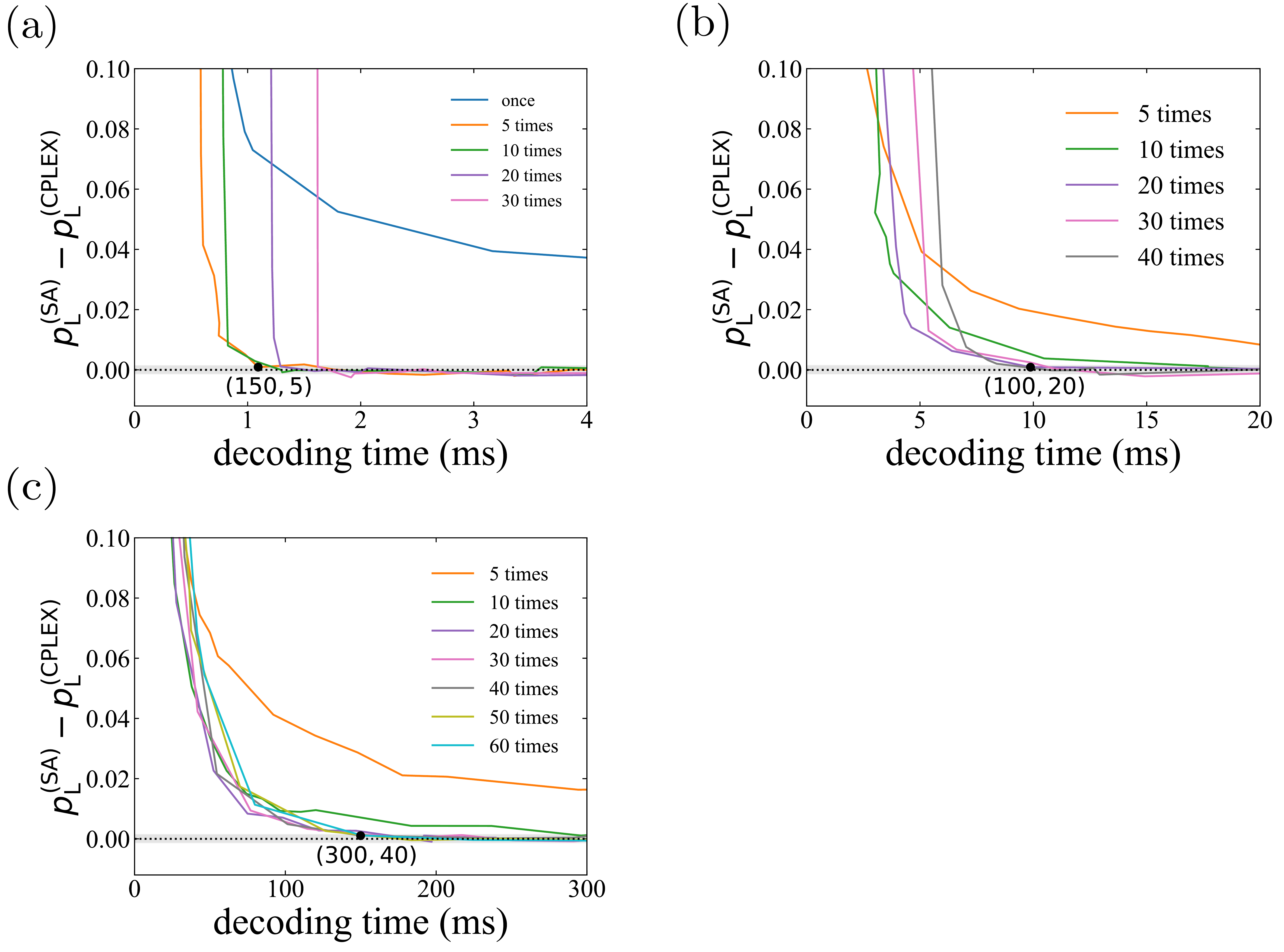}
  \end{center}
  \caption{\textbf{Dependence of $p^{(\mathrm{SA})}_{\mathrm{L}}-p^{(\mathrm{CPLEX})}_{\mathrm{L}}$ on annealing parameters in the phenomenological noise model.} (a) $d=3$. (b) $d=5$. (c) $d=7$. The physical error rate assumed in this figure is 2.80\%.}
  \label{fig:parameter_phenomeno}
\end{figure}

\newpage
Around the threshold error rate, the change of the difference between $p^{(\mathrm{SA})}_{\mathrm{L}}$ and $p^{(\mathrm{CPLEX})}_{\mathrm{L}}$ with respect to the calculation time with varying temperature cycles while keeping the number of iterations fixed is shown in Figs.~\ref{fig:parameter_bitflip}-\ref{fig:parameter_phenomeno}.
The number of Monte Carlo simulation samples in these figures is $5 \times 10^4$.
The legend in the figures represents the number of iterations. The gray band indicates the range of the error bar of $p^{(\mathrm{CPLEX})}_{\mathrm{L}}$. The black circle indicates the point where $p^{(\mathrm{SA})}_{\mathrm{L}}$ reaches the range of the error bar of $p^{(\mathrm{CPLEX})}_{\mathrm{L}}$ and has the shortest calculation time.
Also, the corresponding number of temperature cycles and iterations for that point are shown in the figures in the format of (the number of temperature cycles, the number of iterations).

When the number of iterations is too low, the rate of decrease in the logical error rate with respect to an increase in the number of temperature cycles becomes significantly degraded, leading to a very long time required for accurate decoding. 
On the other hand, if the number of iterations is increased too much, the impact of the overhead required in the annealing process other than state updates will become larger, and the time required for accurate decoding will become longer. 
Therefore, the annealing parameter that achieves the shortest calculation time with high accuracy is realized with an appropriate number of iterations.

We should note that the parameters were optimized by estimating the computation time assuming that the iterations are not parallelized. The best annealing parameters will change if parallel computation is available. Regarding the iterations, simple parallelization can accelerate with 100\% efficiency, so decoding time can be reduced by at least one to two orders of magnitude.

\bibliography{main}
\end{document}